\documentclass[preprint]{revtex4-1}
\usepackage{graphicx}

\begin{document}
\title{Observation of vibrational overtones by single molecule resonant photodissociation}

\author{Ncamiso B. Khanyile}
\author{Gang Shu}
\author{Kenneth R. Brown}
\email{ken.brown@chemistry.gatech.edu}
\affiliation{Schools of Chemistry and Biochemistry, Computational Science and Engineering, and Physics, Georgia Institute of Technology, Atlanta, Georgia, 30332-0400, USA }

\begin{abstract}
Molecular ions can be held in a chain of laser-cooled atomic ions by sympathetic cooling. This system is ideal for performing high-precision molecular spectroscopy with applications in astrochemistry and fundamental physics. Here we show that this same system can be coupled with a broadband laser to discover new molecular transitions. We use three-ion chains of Ca$^{+}$ and CaH$^{+}$ to observe vibrational transitions via resonance enhanced multiphoton dissociation detected by Ca$^{+}$ fluorescence. Based on theoretical calculations, we assign the observed peaks to the transition from the ground vibrational state, $\nu=0$, to $\nu=9$ and $\nu=10$. Our method allows us to track single molecular events, and it can be extended to work with any molecule by using normal mode frequency shifts to detect the dissociation. This survey spectroscopy serves as a bridge to the precision spectroscopy required for molecular ion control.
\end{abstract}

\maketitle

\section*{Introduction}
Precision spectroscopy of molecules and molecular ions can yield insight into the fundamental physical constants and astrochemical processes \cite{Carr2009, Dulieu2011}. Coulomb crystals composed of laser-cooled atomic ions and molecular ions provide a pristine environment for studying the properties of molecules \cite{Willitsch2012}. The laser-cooled atomic ions serve as both a coolant that reduces the temperature and a sensitive detector that allows for single molecule measurements.  The system has been used to study fundamental reaction dynamics ranging from kinetic isotope effects \cite{Staanum2008} to observing the relative reaction rates of molecules whose shape differs by the orientation of a single bond \cite{Chang2013}. It is also a natural system for precision measurements of molecular ion transitions. Molecular transitions allow for the precise study of fundamental constants using physics that is inaccessible in atomic systems \cite{Schiller2005, Demille2008, Baron2014}. Spectroscopy with Coulomb crystals has been used to directly measure dipole-forbidden transitions in N$_2^{+}$ \cite{Germann2014} and accurately determine the rovibrational spectrum of HD$^+$ \cite{Bressel2012}.  Future experiments to test the stability of fundamental constants in time have been proposed \cite{Schiller2005} using the two-ion-species techniques already exploited in the most precise atomic ion clocks \cite{Schmidt2005a, Rosenband2008}.  

Expanding these techniques to a wider array of molecular ion species remains a challenge due to the lack of experimental data on molecular ions transitions. This requires new methods for obtaining spectral information. The spatial localization of molecular ions in a Coulomb crystal results in the required ion density for spectroscopy with low ion numbers relative to traditional techniques. Systems built for high-precision measurement are often incompatible with the survey spectroscopy required to find unknown transitions. However, this is not the case for Coulomb crystals, where the long ion storage time provides multiple opportunities to probe the molecule and the fluorescence of the laser-cooled atomic ion serves as a fast, low-noise detector. CaH$^+$ is a candidate molecule for testing the possible variation in the proton-to-electron mass ratio \cite{Kajita2009}. It is also expected to be relatively abundant in space due to observation of CaH, but has not yet been directly observed \cite{Canuto1993}. In both cases, laboratory measurements of rovibrational transitions in CaH$^+$ are required for scientific progress.

Resonance Enhanced Multi-Photon Dissociation~(REMPD) is a common tool in physical chemistry where conditional on a resonant transition a final photon dissociates the molecule. It can be applied to study the vibrations of molecules from diatomics with a single electron \cite{Roth2006} to peptides \cite{Rizzo2009}. Ion fragments are typically detected by an electron multiplier after mass selection \cite{Rizzo2009, Schneider2014}. Embedding the molecular ions in laser-cooled atomic ions opens up other approaches of detection based on changes in the Coulomb crystal shape \cite{Tong2010} and modulation of the atomic ion fluorescence by driving the resonant motion of the molecular ions \cite{Roth2006}. Resolved sideband spectroscopy on a narrow transition of the atomic ions can also be used to identify the fragments \cite{Goeders2013}.  For the case of CaH$^{+}$, the measurement is simplified by the trapped fragment Ca$^{+}$ fluorescing brightly. This property was used in an ion recycling approach to measure the reaction kinetics of Ca$^{+}$ + HD by direct dissociation \cite{Hansen2012}.  

We present the observation of two vibrational overtones of CaH$^+$ by two photon resonant photodissociation of single molecular ions sympathetically cooled by two Ca+ ions. The overtone is excited by a mode-locked laser in order to span the thermally occupied rotational states.  A second photon dissociates the molecule and the resulting Ca$^+$ is detected by a change in fluorescence.  The change in the rate of dissociation as a function of the mode-locked laser wavelength reveals two peaks.  We assign these peaks to the overtone transtions from the ground vibrational state $\nu=0$ to $\nu=9$ and $\nu=10$ based on theoretical predictions. Our observation is the first measurement of a rovibrational transition in CaH$^+$ and shows that single molecular ion experiments can be used  to discover new transitions.

\section*{Results}
 
\subsection*{Trapping and sympathetically cooling CaH$^+$}

We start with loading three $^{40}$Ca$^{+}$ ions by isotope selective photoionization and then laser cooling the ions to form a three ion chain. The $^{40}$Ca$^{+}$ ions are trapped in a rf Paul trap (r$_{0} = 0.5$~mm) driven at $\Omega =  2\pi\times14$~MHz to confine the ions radially, while static DC voltage applied at the endcaps  confines the ions axially. The trap is kept at a base pressure of about  $4\times10^{-9}$~Pa. The ions are detected by laser induced fluorescence at 397~nm onto a photon multiplier tube (PMT) and electron multiplying charge coupled device (EMCCD) camera. A narrow bandpass filter is used to ensure that only 397nm light is detected. The experimental apparatus has previously been employed for sympathetic heating spectroscopy \cite{Clark2010} and the sideband cooling of molecular ions \cite{Rugango2015}

A $^{40}$CaH$^{+}$ molecule is produced by leaking about $5\times10^{-7}$ Pa of molecular H$_{2}$ into the chamber via a leak valve. The  $^{40}$CaH$^{+}$ is produced via reactive collisions in the gas phase between $^{40}$Ca$^{+}$(4{\em P}$_{\frac{1}{2}}$) and the H$_{2}$ as $^{40}$Ca$^{+}$ + H$_{2}$ $\rightarrow$ $^{40}$CaH$^{+}$ + H.  The product molecule quickly relaxes to its electronic ground state X$^{1}\Sigma^{+}$ (Fig. 1) and the motion of the molecule is sympathetically cooled by the two Ca$^+$ ions. The occurrence of a  reaction is determined when one of the ions goes dark and there is a drop in fluorescence counts. Once a reaction occurs, the leak valve is closed and the experiment is delayed until the base pressure is reached.  The identity of the molecule can be determined by resolved sideband spectroscopy \cite{Goeders2013} and under these experimental conditions we have only observed the formation of CaH$^+$.

A pinhole before the PMT allows partial light collection from all three ions and reduces background due to scattered light. Misalignment from the crystal center results in three distinct collection efficiencies for each ion position. This allows us to detect the position of the dark ion from the fluorescence as shown in Fig. 2. The ion position shifts are due to collision with background gas. Small ion chains of any length could be used to collect the data and test experiments were performed loading two to four Ca$^+$ ions. Three ion chains were preferred due to faster molecule formation time relative to two ion chains and improved stability relative to four ion chains.

\subsection*{Population of internal states of CaH$^+$}

Although the molecular ion is at a translational temperature of a few millikelvin, the internal degrees of freedom are in equilibrium with the room temperature vacuum chamber via black body radiation. A simplified energy level diagram of CaH$^+$ is shown in Fig. 1.  The calculated vibrational frequency of the molecule is $1478.4$~cm$^{-1}$ and we expect that the molecule will be in the ground vibrational state  X$^{1}\Sigma^{+}$ greater than $99.9\%$ of the time \cite{Abe2010}.   On the other hand, the calculated ground state rotational constant, $4.711$~cm$^{-1}$, is small relative to room temperature. The rotational states will be populated with an expected value of {\em J} = $5.36$ and the lowest ten {\em J} states are expected to have more than $94\%$ of the population. Our experiment uses a single molecule at a time, but the blackbody radiation will randomize the {\em J} state on the order of minutes.  To cover this rotational broadening we use a $150$~fs mode-locked  laser to drive the overtones. Alternative approaches include reducing the rotational state space by sympathetic cooling with a buffer gas \cite{Hansen2014} or a cloud of laser-cooled neutral atoms \cite{Rellergert2013} or prepare the molecular ions at a specific state by threshold ionization \cite{Tong2010}. Rotational cooling by optical pumping \cite{Staanum2010, Schneider2010, Lien2014} is currently impossible for CaH$^{+}$ due to the lack of data on bound transitions.

\subsection*{Single CaH$^+$ REMPD}

\begin{figure*}
\centering
\includegraphics[width=\textwidth]{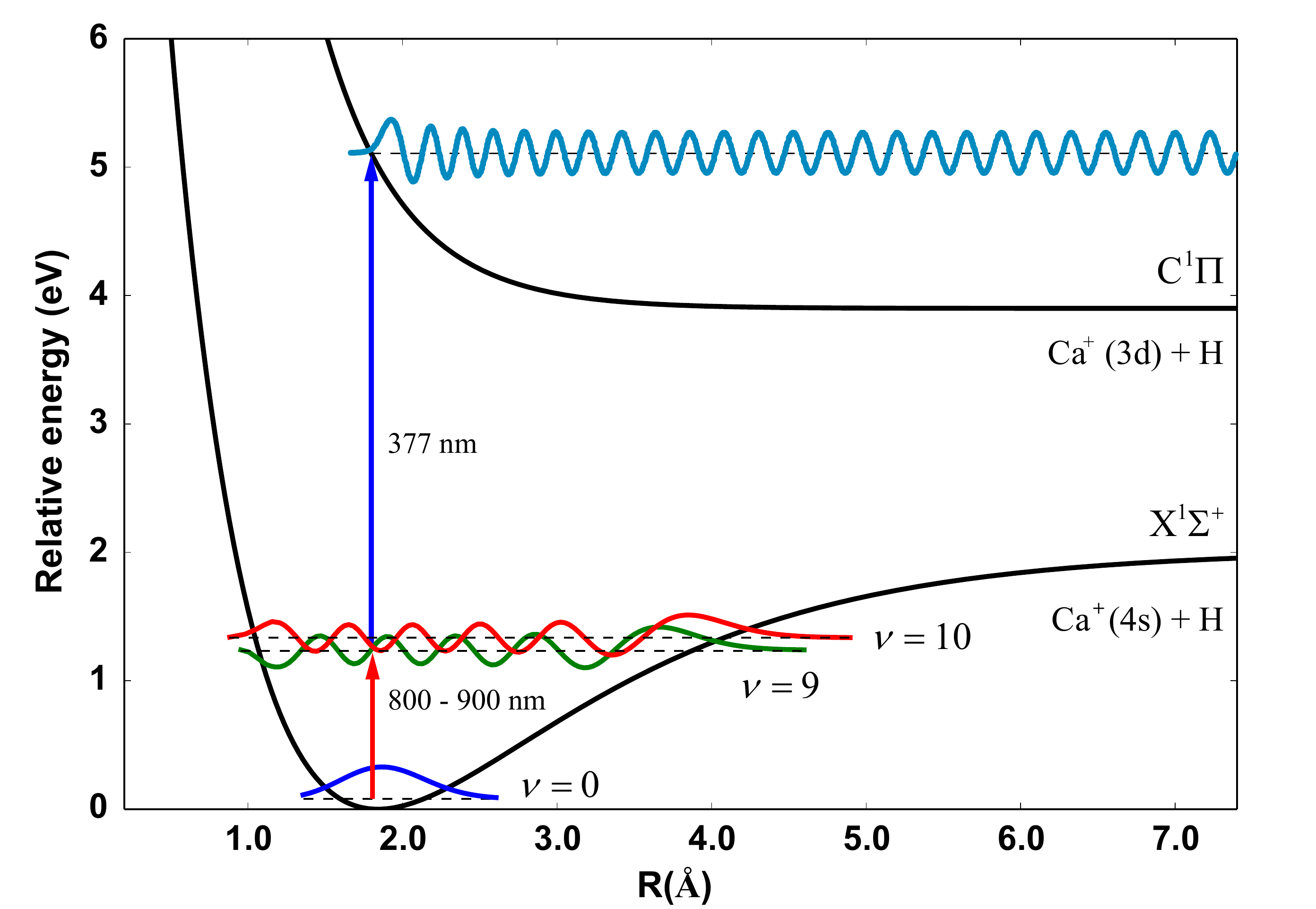}
\caption{{\bf Energy level diagram of CaH$^+$.} Simplified CaH$^+$ energy level diagram showing the overtones excited by a pulsed, tunable infrared laser (800-900 nm).   A second ultraviolet laser (377 nm) excites the overtones to the unbound state to dissociate the molecule.}
\label{fig:EnergyLevels}
\end{figure*}

The two photon photodissociation occurs by first exciting a vibrational overtone of X$^{1}\Sigma^{+}$ and then photodissociating to the state C$^{1}\Pi$  (Fig.~\ref{fig:EnergyLevels}).  We drive the overtone using of a femtosecond mode-locked spectroscopy laser with $\approx 5$~nm bandwidth allowing us to ignore the line broadening due to rotational transitions. The second photon comes from a continuous-wave laser at $377$~nm. The powers of the two lasers are 800 mW and 200 $\mu$W respectively. Upon dissociation, the previously dark  $^{40}$CaH$^{+}$ ion, will be broken into Ca$^{+}$ + H, the Ca$^{+}$ will fluoresce again and there will be an increase in fluorescence counts as shown in Fig. 2.  For each central wavelength of the spectroscopy laser, we expose the molecular ion to the dissociation lasers and record the time for the molecule to dissociate, $\tau_{d}$. The transition strength is proportional to the rate $r_d= 1/\langle \tau_{d}\rangle$.  We repeat the experiment eight times for each wavelength, which resulted in sufficient signal to noise for this survey spectroscopy.   Control experiments with the mode-locked laser blocked were used to rule out other processes, including two UV photon absorption and UV induced electron bombardment.

The spectrum clearly shows two peaks which we identify as the $\nu'=10 \leftarrow \nu=0$ and $\nu'=9 \leftarrow \nu=0 $ overtones of CaH$^{+}$  based on theoretical calculations using NRel/cc-pCV5Z/CASPT2 \cite{Abe2010}~(Fig.~\ref{fig:Spectra}). We measure the $\nu'=10 \leftarrow \nu=0$ transition to be at 812(3) nm compared to the theoretical value of  813.3 nm and $\nu'=9 \leftarrow \nu=0 $ transition to be at 890(3) nm  compared to the theoretical value of 883.3 nm. This disagreement is in line with observed differences between calculated and measured vibrational transition frequencies in other metal hydrides \cite{Abe2010}.

\section*{Discussion}
Our experimental setup was intended for high-precision quantum logic spectroscopy \cite{Schmidt2005a} experiments on molecular ions. We have shown that the same setup can be used for the preliminary large range spectroscopy necessary to observe even weak lines despite trapping only a few ions at a time. The next step for precision spectroscopy of CaH$^+$  is to reduce the rotational temperature by sympathetic cooling with neutral atoms and then rotationally resolve these transitions and the fundamental transition \cite{Rellergert2013, Hansen2014}. Then quantum logic spectroscopy can be performed on ground state cooled Ca$^+$-CaH$^+$ crystals \cite{Rugango2015} in order to reach the precision necessary for observing relative changes in fundamental constants \cite{Rosenband2008}. 

Our current experiment takes advantage of the fluorescence of the dissociated product and can be easily applied to other fluorescent product ions. The method can be modified for dark product ions by periodically blocking the dissociation laser and measuring the mass of the product ion. This can be achieved by monitoring the sidebands of the atomic ions and calculating the mass of the unknown ion from the observed normal modes \cite{Goeders2013}. The sideband monitoring can be used to extend few ion crystal spectroscopy beyond REMPD to any action spectroscopy where the laser-cooled atomic ion is inert. 

We expect that few ion Coulomb crystals will provide a method for discovering the spectra of ions that are difficult to make in the high numbers required for other spectroscopic methods. As a point of comparison consider the recently reported results on the spectroscopy of CH$_5^+$ \cite{Asvany2015}. The spectra requires trapping a few thousand molecular ions per frequency. At equilibrium, approximately ten percent of those ions form complexes with the He buffer gas, as detected by mass-spectrometry. The number of complexes is partially depleted when an internal transition is excited through laser induced inhibition of complex growth (LIICG) \cite{Chakrabarty2013}. In the few ion crystal experiment, a single CH$_5^+$ would be trapped and the normal modes of the crystal monitored. The formation of the CH$_5^+ \cdot$He would result in a change in the normal mode frequencies. The system is in dynamic equilibrium and for every applied laser frequency the normal modes are measured multiple times.  The spectral signal is the fraction of measurements where CH$_5^+ \cdot$He is observed. When the excitation laser is on resonance, the signal is suppressed.  In principle, one could obtain the entire LIICG spectra of CH$_5^+$ using only a single molecular ion.   In practice, the rates of trapping molecules, forming complexes, and detecting normal modes will determine when the single molecule approach is preferable to other methods.

\begin{description}
\item[Acknowledgements] We thank M. Abe, S. Janardan, M. Kajita, and E. Hudson for useful discussions, Y. Choi for assistance with data collection, and D. Denison for use of the MIRA. This work was supported by ARO grant W911-NF-12-1-0230.
\item [Author Contributions] K.R.B. and N.K. conceived the experiment and wrote the manuscript. N.K. and G.S. collected the data. K.R.B., N.K., and G.S. performed the data analysis and prepared the figures.
\item [Author Information] The authors declare no competing financial interests. Correspondence and requests for materials should be addressed to K.R.B. (ken.brown@chemistry.gatech.edu)
\end{description}

\newpage

\begin{figure*}
\centering
\includegraphics[width=\textwidth]{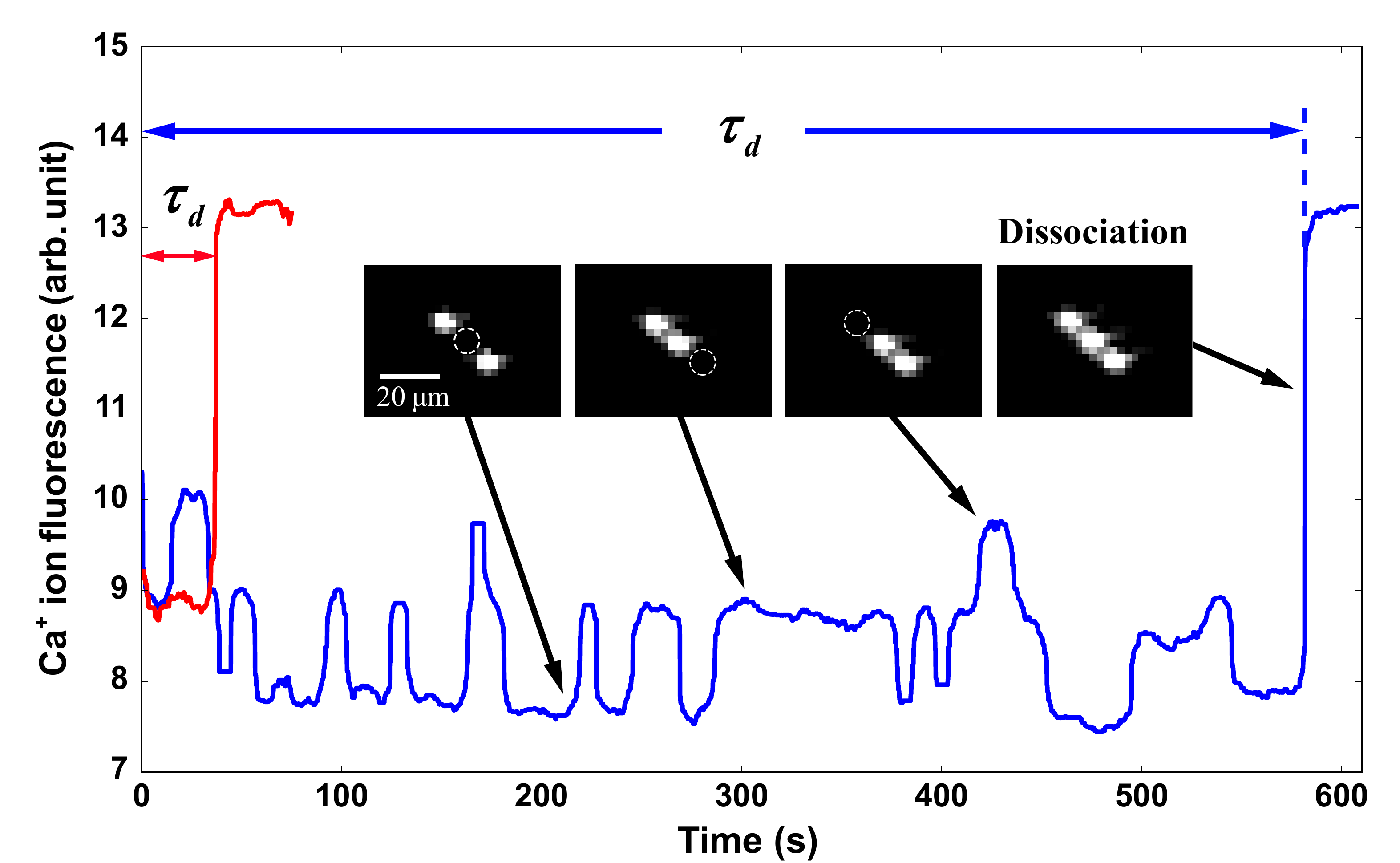}
\caption{{\bf Dissociation measurement.} After reaction with hydrogen, a mixed Coulomb crystal of two Ca$^+$ ions and one CaH$^+$ ion. The dissociating lasers are applied and the time to dissociation, $\tau_d$, is measured by observing the change in fluorescence. The red trace shows an  example dissociation event when the infrared laser is resonant with an overtone. The blue trace is an event when the infrared laser is blocked. A pinhole in the optical system allows us to correlate fluorescence with not only the number of Ca$^+$ ions, but also the relative position of the molecular ion. }
\label{fig:ExampleTrace}
\end{figure*}

\begin{figure*}
\centering
\includegraphics[width=\textwidth]{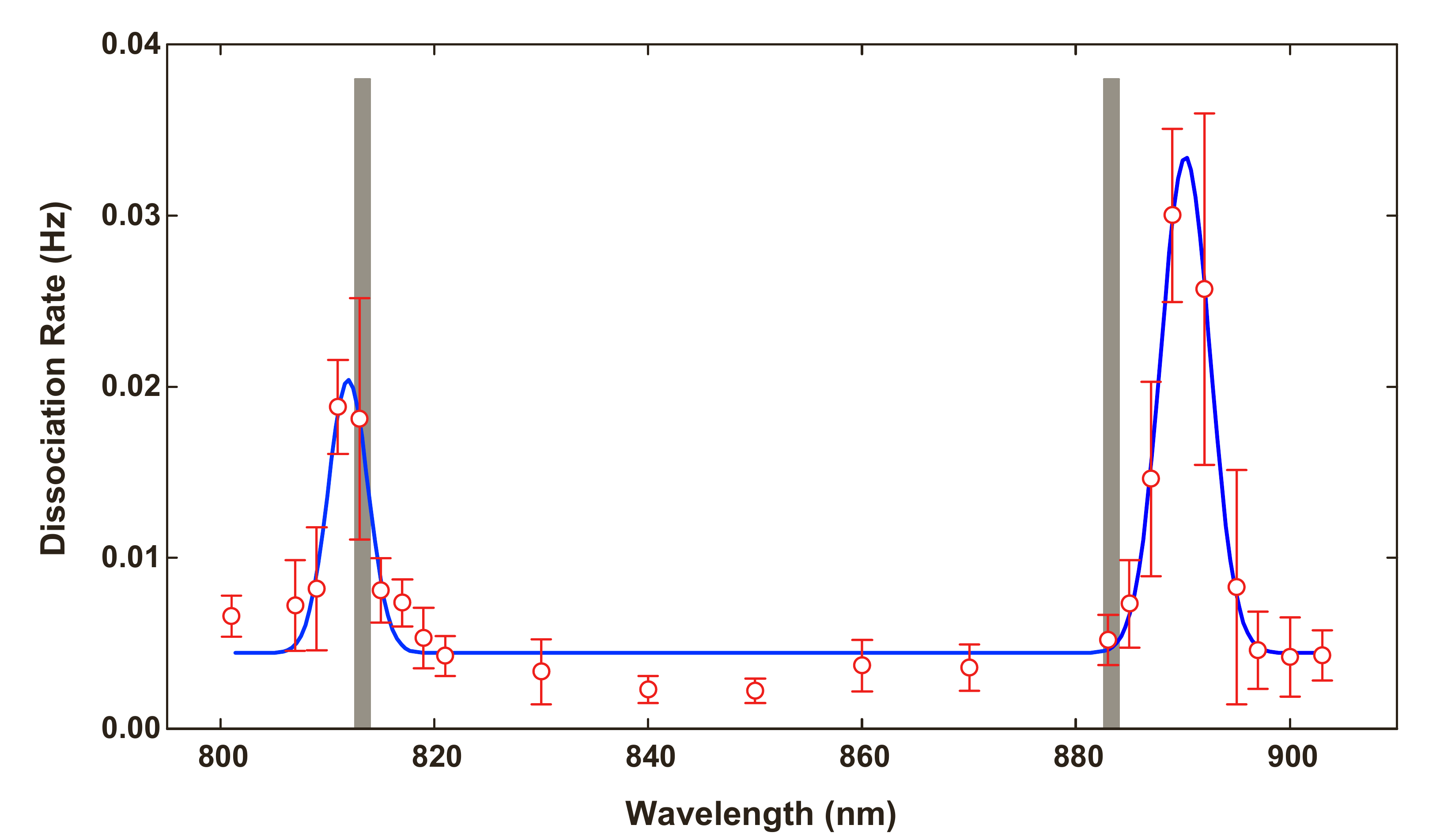}
\caption{{\bf CaH$^+$ vibrational overtone spectra.} The measured $\tau_d$ are averaged over eight experiments and the inverse is plotted as a function of the IR wavelength.  The data reveals two peaks which are fit assuming a Gaussian line shape. Gray bars are centered at the calculated theoretical values \cite{Abe2010} for the $\nu'= 10 \leftarrow \nu=0$ and $\nu'=9 \leftarrow \nu=0 $ overtones. Error bars are the standard error of the mean of $r_d=1/\langle \tau_d \rangle$, $\epsilon_r$,  propagated from the standard error of the mean of $\langle \tau_d \rangle$, $\epsilon_\tau$: $\epsilon_r=r_d^2\epsilon_\tau$.  }
\label{fig:Spectra}
\end{figure*}

\end{document}